\documentclass[conference]{IEEEtran}

\usepackage[letterpaper, left=1in, right=1in, bottom=1in, top=0.75in]{geometry}
\usepackage{epsfig,amsmath,amssymb,amsfonts}
\usepackage{wasysym}
\usepackage{algorithmic}
\usepackage{algorithm}
\usepackage{epstopdf}
\usepackage[acronym]{glossaries}
\usepackage{cite}
\usepackage{amsthm}
\usepackage{enumerate}
\usepackage{cases}
\usepackage{subfigure}
\usepackage{graphicx}

\IEEEoverridecommandlockouts \IEEEpubid{\makebox[\columnwidth]{ 978-1-5386-3531-5/17/\$31.00~\copyright~2017 IEEE \hfill} \hspace{\columnsep}\makebox[\columnwidth]{ }}

\begin{document}
\title{Use of Two-Mode Circuitry and Optimal Energy-Efficient Power Control Under Target Delay-Outage Constraints}

\author{\IEEEauthorblockN{Jinkun Xu, Yu Chen, Hao Chen, Qimei Cui and Xiaofeng Tao}
\IEEEauthorblockA{National Engineering Laboratory for Mobile
Network Technologies\\Beijing University of Posts and Telecommunications\\
Email: \{yu.chen, cuiqimei\}@bupt.edu.cn}
}

\maketitle

\begin{abstract}
An accurate energy efficiency analytical model based on a two-mode circuitry was recently proposed; and the model showed that the use of this circuitry can significantly improve a system's energy efficiency. In this paper, we use this analytical model to develop a new power control scheme, a scheme that is capable of allocating a minimum transmission power precisely within the delay-outage probability constraint. Precision brings substantial benefits as numerical results show that the energy efficiency using our scheme is much higher than other schemes. Results further suggest that data rate values affect energy efficiency non-uniformly, i.e., there exists a specific data rate value that achieves maximum energy efficiency.
\end{abstract}

\IEEEpeerreviewmaketitle

\section{Introduction}
The energy efficiency and the end-to-end latency in 5G systems are expected to have a 100-fold increase and a 10-fold reduction over 4G systems, respectively~\cite{ref1}. Energy efficiency is a physical-layer measure that is often related to power control problems while delay is a link-layer measure. Therefore, it is important to design a cross-layer power control scheme that improves energy efficiency as well as takes delay QoS requirements into account.

If a system's energy efficiency is measured in bits-per-Joule, then the earliest study of this topic may date back to 2012 when Musavian and Le-Ngoc~\cite{ref7}, \cite{ref8} defined a cross-layer energy efficiency as the ratio of the effective capacity to the total power consumption. Based on this measure, they then proposed a power allocation scheme to maximize energy efficiency under delay-outage constraints. Cheng et al.~\cite{ref9} developed statistical delay-bounded QoS-driven energy-efficient power allocation schemes over SISO and MIMO-based wireless systems. Zhao and Wang~\cite{ref10} considered a multi-user massive MIMO system and proposed an energy-efficient power allocation scheme while guaranteeing the delay outage requirements of each user. Although the effective capacity model describes a delay-outage probability on a predefined delay bound by two effective capacity functions (the \emph{nonempty buffer probability} and the \emph{QoS exponent}~\cite{ref2}), all the above work only considered one of them--the QoS exponent, resulting in overestimated delay-outage probability values.

In 2016, Sinaie et al.~\cite{ref11} developed the first version of a power control scheme that considers 1) the use of two-mode circuitry for better energy efficiency and 2) the use of both two effective capacity functions for better delay-outage provisions. By continuing Sinaie's work, a more accurate energy efficiency analytical model based on a two-mode circuitry was recently proposed by us~\cite{ref12}. On the other hand, the nonempty buffer probability in Sinaie's work was approximated by the ratio of the average arrival rate to the average service rate, which is less accurate than the method proposed by Chen and Darwazeh~\cite{ref14}.

By assuming a two-mode transceiver circuitry is in use and the transmission power is constant over time, we will in this paper find an optimal power control scheme that maximizes a system's energy efficiency as well as precisely guarantees a target delay-outage probability. Specifically, a new accurate delay-outage probability approximation method is first developed by following Chen and Darwazeh's work. This method together with our energy efficiency analytical model~\cite{ref12} will then be used to develop our power control scheme.

The remainder of this paper is organized as follows. Section II describes the wireless communication system model and the cross-layer energy efficiency using a two-mode transceiver circuitry. In section III, we formulate and solve the energy-efficient power control problem under target delay-outage constraints. In section IV, we compare the simulation and numerical results with approximation results using different power control schemes. Section V summarizes our work.

\section{System Model}

A point-to-point wireless communication system over a block-fading channel is considered, as shown in Fig. 1. It contains five basic components: data source, transmitter, wireless channel, receiver and data sink. The system is discrete in time; the duration of a slot, denoted by $T_s$, is assumed to be equal to the length of a fading block~\cite{ref13}.

The data source block generates data and feeds them into a buffer with infinite buffer size. The arrival data length from this block at slot $n$ is $A[n]~(n=\{1,2,...\})$. We assume that
\begin{enumerate}
\item the data arrivals confirms to a Bernoulli process with a data arrival probability $p\ (p\ \in(0,1])$ and
\item if data arrives at one slot, its length is exponentially distributed with an average data length $\overline{L}$.
\end{enumerate}
Based on the above assumptions, the arrivals $A[1],A[2],...$ are independent and identically distributed (IID) random variables (RVs) identical to a RV $A$; and the probability density function (PDF) of the arrival $A$ is \cite{ref12}
\begin{equation}
{{f_A}(a)=}
\begin{cases}
p\frac{1}{\overline{L}}\exp ( - \frac{1}{\overline{L}}a) & (a > 0),\\
1 - p & (a = 0).
\end{cases}
\label{eq:1}
%(1)
\end{equation}
The average arrival rate $\mu$ is therefore
\begin{equation}
\mu = \overline{L}p/T_s.
\label{eq:2}
%(2)
\end{equation}

Denote by $S[n]$ the amount of data the transmitter is capable of transmitting at slot $n$. We assume that the services $S[1] ,S[2],...$ are IID RVs identical to a RV $S$ with any type of probability density functions $f_S(s)$.

\subsection{Use of Two-Mode Transceiver Circuitry and Its Cross-Layer Energy Efficiency Definition}

In the system model shown in Fig. 1, both the transmitter and the receiver use a two-mode circuitry, a circuitry that works either in the transmission mode if there is data to transmit or in the idle mode otherwise. The total power consumption in these two modes is different and can be expressed as follows:
\begin{equation}
{P_{tx{\rm{\_}}mode}} = {P_c} + {P_{tx}}
\label{eq:3}
%(3)
\end{equation}
and
\begin{equation}
{P_{idle{\rm{\_}}mode}} = {P_c} + {P_{idle}},
\label{eq:4}
%(4)
\end{equation}
where $P_c$ is the constant circuit power consumption, $P_{tx}$ and $P_{idle}$ are the power consumption of being in the transmission mode and idle mode, respectively. Based on this circuitry model, the cross-layer energy efficiency has been found in~\cite{ref12}:
\begin{equation}
\begin{aligned}
\eta &=\frac{\mu}{{{P_{tx}}{p_{tx}}+{P_{idle}}{p_{idle}}+{P_c}}}\\
     &=\frac{\mu}{{{P_{tx}}+{P_c}-({P_{tx}}-{P_{idle}}){u^*}\overline L(1-p)}},
\label{eq:5}
%(5)
\end{aligned}
\end{equation}
where $p_{tx}$ and $p_{idle}$ are the transmission probability and the idle probability, and $u^*>0$ is termed the QoS exponent and it is a unique solution of the equation (eq. (27) in \cite{ref12}):
\begin{equation}
\frac{{1 - u\overline L }}{{1 - \left( {1 - p} \right)u\overline L }} =
\int_0^\infty  {{e^{ - us}}} {f_S}\left( s \right)ds.
\label{eq:6}
%(6)
\end{equation}

\subsection{Power Control under Target Delay-Outage Constraints}

In order to support QoS, the transmitter determines the transmission power $P_{tx}$ based on the target delay-outage probability constraint (required by upper-layer traffic flows) and the channel state information (CSI) fed back from the receiver~\cite{ref4}. When a maximum delay bound $D_{max}$ and a tolerance $\varepsilon$ are predefined, the system is in delay-outage if it cannot guarantee the following inequality:
\begin{equation}
{\rm P}\left( {D > {D_{\max }}} \right) \le \varepsilon.
\label{eq:7}
%(7)
\end{equation}
In this respect, (\ref{eq:7}) is termed the delay-outage probability; the target delay-outage constraint for a traffic flow can be specified by the pair $\{ {D_{\max }},\varepsilon \} $.

\begin{figure}[!t]
\centering
\includegraphics[width=2.9in]{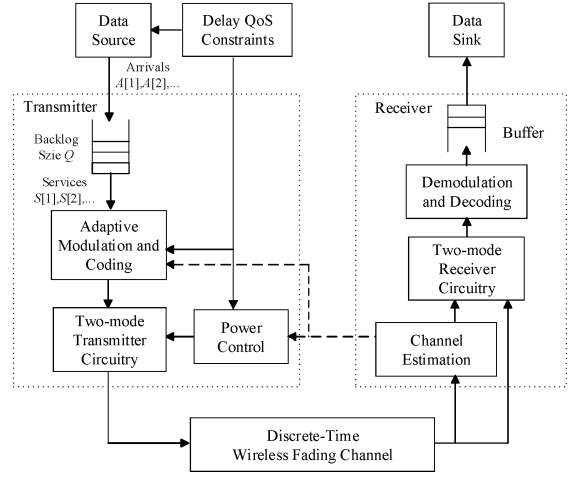}
\caption{System model.}
\end{figure}

\section{Energy-Efficient Power Control Under Target Delay-Outage Constraints}

When a new constraint---the target delay-outage constraint $\{ {D_{\max }},\varepsilon \} $ is introduced to our system model, the optimal power control problem under this constraint can be expressed based on (\ref{eq:5})--(\ref{eq:7}):
{\small
\begin{equation}
\textbf{P1:}\quad\max\eta\!=\!\max\!\frac{\mu}{{{P_{tx}}\!+\!{P_c}\!-\!({P_{tx}}\!-\!{P_{idle}}){u^*}\!\overline L\!(1\!-\!p)}},
\label{eq:8}
%(8)
\end{equation}
}
\begin{equation}
s.t. {\rm P}\left( {D > {D_{\max }}} \right) \le \varepsilon,\qquad\qquad\qquad\quad
\label{eq:9}
%(9)
\end{equation}
\begin{equation}
\frac{{1 - {u^*}\overline L }}{{1 - \left( {1 - p} \right){u^*}\overline L }} =
\int_0^\infty  {{e^{ - {u^*}s}}} {f_S}\left( s \right)ds,\!\!\!
\label{eq:10}
%(10)
\end{equation}
\begin{equation}
0 < p < 1,\qquad\qquad\qquad\qquad\qquad\quad\;\;
\label{eq:12}
%(12)
\end{equation}
\begin{equation}
\overline{L} > 0.\qquad\qquad\qquad\qquad\qquad\qquad\quad
\label{eq:13}
%(13)
\end{equation}
The third and fourth constraints (expressed in (\ref{eq:12}) and (\ref{eq:13}) respectively) are implicit so will be omitted in the rest of this work. The delay-outage probability constraint (\ref{eq:9}) is still unknown at the moment but will be explained in terms of transmission power in Section III-A. A solution to the problem \textbf{P1} is given in Section III-B.

\subsection{Delay-Outage Probability Approximation Method}
By following Chen and Darwazeh's work \cite{ref14}, the delay distribution can be obtained in the proposition below:

\emph{Proposition 1:} Consider a system that has IID services, a Bernoulli arrival process with data length being exponentially distributed. If there exists a unique $u^*$ that satisfies the equation (\ref{eq:6}), the complementary cumulative distribution function (CCDF) of delay can be approximated by
\begin{equation}
{\rm P}\left( {D > t} \right) \approx {p_w}{\left( {\frac{{1 - {u^*}\overline L}}{{1 - {u^*}\overline L + p{u^*}\overline L}}} \right)^{\frac{t}{{{T_s}}}}},
\label{eq:14}
%(14)
\end{equation}
where
\begin{equation}
p_w = {\rm P}(D>0) = \frac{{{1 - u^*\overline L}}}{{1 - {u^*}\overline L + p{u^*}\overline L}}.
\label{eq:14_1}
\end{equation}

For a proof of Proposition 1, see Appendix A. ${p_w}$ is termed the \emph{nonzero delay probability}. Proposition 1 indicates that a predefined target delay-outage constraint $\{ {D_{\max }},\varepsilon \} $ can be expanded as
{\small
\begin{equation}
{\rm P}\left( {D > {D_{\max }}} \right) \approx {\left( {\frac{{1 - {u^*}\overline
L}}{{1 - {u^*}\overline L + p{u^*}\overline L}}} \right)^{\frac{{{D_{\max }}}}{{{T_s}}} +
1}} \le \varepsilon.
\label{eq:15}
%(15)
\end{equation}
}

In the next step, we ought to find a relation between the transmission power $P_{tx}$ in (\ref{eq:8}) and the QoS exponent $u^*$ in (\ref{eq:14}). Denote by $C[n]$ the service rate at slot $n$. Over a block-fading channel, $C[n]$ and $S[n]$ are linearly related:
\begin{equation}
S\left[ n \right] = {T_s}C\left[ n \right].
\label{eq:16}
%(16)
\end{equation}
Based on Shannon's capacity equation and the assumption that the knowledge of CSI is perfectly known at the receiver side, the service rate at slot $n$ is bounded by \cite{ref7}
\begin{equation}
C[n] = {B_c}\log_2(1 + \frac{{{P_{tx}}\gamma [n]}}{{{L_p}{N_0}{B_c}}}),
\label{eq:17}
%(17)
\end{equation}
where $B_c$ is a channel bandwidth, $L_p$ is a distance-based path loss, $N_0$ is the noise spectral density and $\gamma{[n]}$ is the single-to-noise ratio (SNR). Moreover, the instantaneous received SNR values $\gamma[1],\gamma[2],...$ are assumed to be IID RVs identical to a RV $\gamma$. The right-hand side of (\ref{eq:10}) can be further expanded by using (\ref{eq:16}) and (\ref{eq:17})  so that (\ref{eq:10}) can be rewritten as
{\small
\begin{equation}
\frac{{1\!-\!{u^*}\overline L }}{{1\!-\!\left({1\!-\!p}\right){u^*}\overline L }}\!=\!\int_0^\infty\!{{e^{\!-\!{u^*}{T_s}{B_c}\log_2(1\!+\!\frac{{{P_{tx}}\gamma
}}{{{L_p}{N_0}{B_c}}})}}} {f_\gamma}\!(\gamma)\!d\gamma\!.
\label{eq:18}
%(18)
\end{equation}
}
Therefore, \textbf{P1} is equivalent to the following problem:
\begin{equation}
\textbf{P2:}\max \frac{\mu }{{{P_{tx}}\!+\!{P_c}\!-\!({P_{tx}}\!-\!{P_{idle}}){u^*}\bar L(1\!-\!p)}},\quad\quad
\label{eq:19}
%19
\end{equation}
\begin{equation}
s.t. {\left( {\frac{{1 - {u^*}\overline L}}{{1 - {u^*}\overline L + p{u^*}\overline L}}}
\right)^{\frac{{{D_{\max }}}}{{{T_s}}} + 1}} \le \varepsilon,\qquad\qquad\quad
\label{eq:20}
%(20)
\end{equation}
{\small
\begin{equation}
\frac{{1\!-\!{u^*}\overline L }}{{1\!-\!\left({1\!-\!p}\right){u^*}\overline L }}\!=\!\int_0^\infty\!{{e^{\!-\!{u^*}{T_s}{B_c}\log_2(1\!+\!\frac{{{P_{tx}}\gamma
}}{{{L_p}{N_0}{B_c}}})}}} {f_\gamma}\!(\gamma)\!d\gamma\!,
\label{eq:21}
%(21)
\end{equation}
}

\subsection{Solution to the Power Control Optimization Problem over Nakagami-$m$ Channels}

The Nakagami-$m$ distribution is a generalized distribution, which can model different fading environments. Given that the PDF of $\gamma$ has a Nakagami-$m$ distribution \cite{ref15}:
\begin{equation}
{f_\gamma }(\gamma ) = \frac{{{\gamma ^{m - 1}}}}{{\Gamma (m)}}{\left(
{\frac{m}{{\overline{\gamma} }}} \right)^m}\exp \left( { - \frac{m}{{\overline{\gamma}
}}\gamma } \right){\rm{ }}\gamma  \ge 0,
\label{eq:23}
%(23)
\end{equation}
where $m$ is the fading parameter, $\Gamma(\cdot)$ is a Gamma function and $\overline{\gamma{}}$ is the average value of SNR. The problem \textbf{P2} can be solved over Nakagami-$m$ channels because of the lemma below:

\emph{Lemma 1:} Consider a wireless system over a Nakagami-$m$ channel, the energy efficiency in \textbf{P2} has an upper bound ${\eta{}}_u$:
\begin{equation}
\begin{aligned}
%{\eta _u} &= \frac{\mu }{{{P_l}{p_{tx}} + {P_{idle}}{p_{idle}} + {P_c}}} \\
%          &= \frac{\mu }{{{P_l} + {P_c} - ({P_l} - {P_{idle}}){u^*}\overline L(1 - p)}},
{\eta _u} &= \frac{\mu }{{{P_l} + {P_c} - ({P_l} - {P_{idle}}){u^*}\bar L(1 - p)}}\\
 &= \frac{\mu}{{{P_l}\left(\!{1\!-\!{u^*}\overline L(1\!-\!p)} \right)\!+\!{P_{idle}}{u^*}\overline L(1\!-\!p)\!+\!{P_c}}},
\end{aligned}
\label{eq:24}
%(24)
\end{equation}
where
\[{P_l} = \frac{{\overline \gamma {L_p}{N_0}}}{{m{T_s}}}\frac{1}{{{u^*}}}\left( {{{\left( {1 + \frac{{p{u^*}\overline L }}{{1 - {u^*}\overline L }}} \right)}^{\frac{1}{m}}} - 1} \right).\]

For a proof of this lemma, see Appendix B. The following proposition plays an important role in solving \textbf{P2}.

\emph{Proposition 2:} Over a Nakagami-$m$ channel, the energy efficiency is a decreasing function of the transmission power $P_{tx}$.

For a proof of Proposition 2, see Appendix C. The statement in Proposition 2 is rather intuitive, i.e., when no other constraints come into force, the best power control strategy is keeping the lowest transmission power possible but still ensures a stable system, i.e.,
\begin{equation}
p\overline L < E\left[ S \right].
\label{eq:25}
%(25)
\end{equation}
The same observations can be found in \cite{ref12} but no proofs are given there. More importantly, the maximization problem of energy efficiency in \textbf{P1} and \textbf{P2} is equivalent to a minimization problem of transmission power, i.e.,
\begin{equation}
\max\eta\!= \min P_{tx}.
\label{eq:25_1}
\end{equation}

The following corollary is an immediate result of Proposition 2.

\emph{Corollary 1:} The solution of \textbf{P2} can be found when the first constraint (\ref{eq:20}) is fulfilled with equality, i.e.,
\begin{equation}
{\left( {\frac{{1 - {u^*}\bar L}}{{1 - {u^*}\bar L + p{u^*}\bar L}}} \right)^{\frac{{{D_{\max }}}}{{{T_s}}} + 1}} = \varepsilon .
\label{eq:26}
%(26)
\end{equation}

\quad\textit{Proof:}
%\begin{proof}
The corollary can be proved by contradiction. Assume ${(u}_1,P_{tx1})$ solves \textbf{P2} and get a maximum energy efficiency ${\eta{}}_1$, but ${(\frac{\left(1-u_1\bar{L}\right)}{\left(1-u_1\bar{L}+pu_1\bar{L}\right)})}^{\frac{D_{max}}{T_s}+1}<\varepsilon$. Because the delay-outage probability is a decreasing function of $u^*$, then there exists $0<u_2<u_1,$ which satisfies ${(\frac{\left(1-u_2\bar{L}\right)}{\left(1-u_2\bar{L}+pu_2\bar{L}\right)})}^{\frac{D_{max}}{T_s}+1}=\varepsilon$ and an energy efficiency ${\eta{}}_2$. Moreover, the energy efficiency is a decreasing function of $u^*$, so ${\eta{}}_2>{\eta{}}_1$. This contradicts the assumption that ${(u}_1,P_{tx1})$ is a solution of \textbf{P2}.
%\end{proof}
\begin{flushright}$\blacksquare$\end{flushright}

\begin{figure}[!t]
\centering
\includegraphics[width=2.9in]{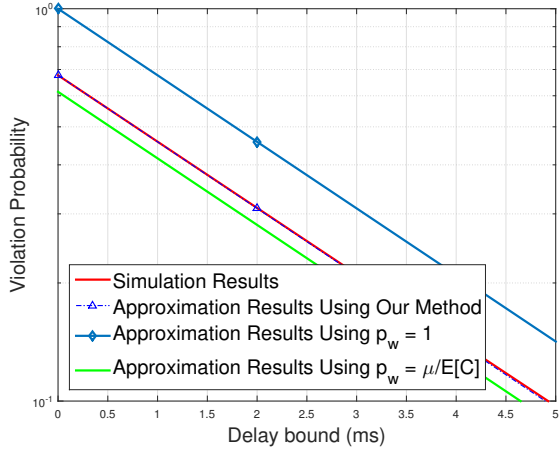}
\caption{Simulation and approximation results of $P(D>t)$ when $\mu=350$ Kbps, $p=0.5$ and $\overline{L}=700$ bits.}
\end{figure}

%\begin{figure}[!t]
%\centering
%\includegraphics[width=2.9in]{fig22.eps}
%\caption{Simulation and approximation results of $P(D>t)$, $\mu=450$ Kbps, $p=0.5$ and $\overline{L}=900$ bits.}
%\end{figure}

Based on Corollary 1, the optimal $u^*$ can be obtained by solving (\ref{eq:26}):
{\small
\begin{equation}
\!\!\begin{aligned}
&{\left( {\frac{{1 - {u^*}\overline L}}{{1 - {u^*}\overline L + p{u^*}\overline L}}}
\right)^{\frac{{{D_{\max }}}}{{{T_s}}} + 1}} = \varepsilon \\
&\Leftrightarrow \frac{{1 - {u^*}\overline L}}{{1 - {u^*}\overline L + p{u^*}\overline L}} =
\frac{1}{\beta } \Leftrightarrow {u^*} = \frac{1}{{\overline L}}\frac{{\beta  - 1}}{{p
+ \beta  - 1}},
\end{aligned}
\label{eq:27}
%(27)
\end{equation}}where $\beta  = {\varepsilon ^{ - \frac{{{T_s}}}{{{D_{\max }} + {T_s}}}}}$. By the result (\ref{eq:25_1}) and substituting $u^*$ in (\ref{eq:20}) with (\ref{eq:27}), we now finalize the power control problem \textbf{P2} as follows:
\begin{equation}
\textbf{P3:}\qquad\qquad\qquad\qquad\min P_{tx}\qquad\qquad\qquad\quad
\label{eq:28}
%(28)
\end{equation}
\begin{equation}
s.t.\frac{1}{\beta }\!=\!\int_0^\infty  {{{\left( {1\!+\!\frac{{{P_{tx}}\gamma
}}{{{L_p}{N_0}{B_c}}}} \right)}^{ - \phi \frac{1}{{\overline L}}\frac{{\beta  - 1}}{{p
+ \beta \!-\!1}}}}} {f_\gamma }\left( \gamma  \right)d\gamma,
\label{eq:29}
%(29)
\end{equation}
where $\phi=\frac{T_sB_c}{\log\left(2\right)}$. The problem \textbf{P3} can be solved when $P_{tx}$  satisfies the constraint (\ref{eq:29}) while this optimal $P_{tx}$ can be obtained numerically via e.g., a binary search algorithm in Appendix D.

\section{Results and Discussion}

We simulate the system model in Section II. An empty buffer is assumed at slot 0. We follow the parameter setting in~\cite{ref7} \cite{ref14} and assume $m$, $\gamma$, $B_c$ and $T_s$ to be 2, 10 dB, 180 KHz and 1 ms, respectively. The path loss for a macro-cell environment with a carrier frequency of 2 GHz is considered~\cite{ref5}:
\[{L_p} = 128.1 + 37.6{\log _{10}}\left( d \right),\]
where $d$ is the distance between the transmitter and the receiver. $P_c$ and $P_{idle}$ are set to 0.1, 0.03 watts, respectively~\cite{ref3}. The rest of the parameters are listed in Table I.

\begin{figure}[!t]
\centering
\includegraphics[width=2.9in]{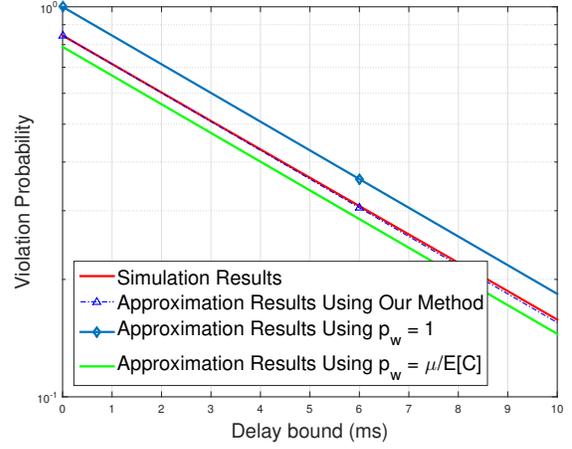}
\caption{Simulation and approximation results of $P(D>t)$, $\mu=450$ Kbps, $p=0.5$ and $\overline{L}=900$ bits.}
\end{figure}

\begin{table} [!tp]
\renewcommand{\arraystretch}{1.3}
\centering
\caption{Simulation Parameters}
\begin{tabular}{r|l}
\hline
Parameter & Value \\
%\hline
%Time duration of a slot, $T_s$ & 1ms \\
%\hline
%System bandwidth, $B_c$ & 180KHz \\
%\hline
%Shape parameter, $m$ & 2 \\
%\hline
%Average SNR, $\gamma$ & 10 dB \\
\hline
Noise spectral density, $N_0$ & -174dBm/Hz \\
\hline
Distance, $d$ & 1 Km \\
\hline
Simulation time & 5,000,000 time slots \\
\hline
Precision tolerance for the  & $10^{-6}$ \\
binary search algorithm, ${\delta}_t$ &  \\
\hline
\end{tabular}
\end{table}

Figs. 2 and 3 show the simulated and approximated CCDF results of delay when the average date rates are 350 Kbps ($p$ = 0.5, $\overline L$ = 700 bits) and 450 Kbps ($p$ = 0.5, $\overline L$ = 900 bits), respectively. The approximation results using our method are shown to be overlapping the simulation results. This indicates that our method is highly accurate and further implies that our power control scheme is capable of providing precise delay-outage probability guarantees.

Two other approximation results using two existing methods are shown in Fig. 2 and 3 as well. For convenience, we reproduce these two methods first:
\begin{enumerate}
	\item assume that $p_w = 1$ and
\begin{equation}
{\rm P}\left( {D > t} \right) \approx {\left( {\frac{{1 - {u^*}\overline L}}{{1 - {u^*}\overline L + p{u^*}\overline L}}} \right)^{\frac{t}{{{T_s}}}}};
\end{equation}
	\item assume that $p_w$ equals the ratio of the average arrival rate to the average service rate $p_{ratio} = \mu/E[C]$ (used in Sinaie's work \cite{ref11}) and
\begin{equation}
{\rm P}\left( {D > t} \right) \approx p_{ratio} {\left( {\frac{{1 - {u^*}\overline L}}{{1 - {u^*}\overline L + p{u^*}\overline L}}} \right)^{\frac{t}{{{T_s}}}}}.
\end{equation}
\end{enumerate}
As shown in these two figures, approximation results using the first method are higher than simulation results so this method \emph{overestimates} delay performance. On the contrary, approximation results using the second method are lower than simulation results so this method \emph{underestimates} delay performance. Any power control based on this method will use less than expected transmission power, which results in a \emph{delay outage}. Thus, such a power control scheme will not be considered in our work.

\begin{figure}[!t]
\centering
\includegraphics[width=2.9in]{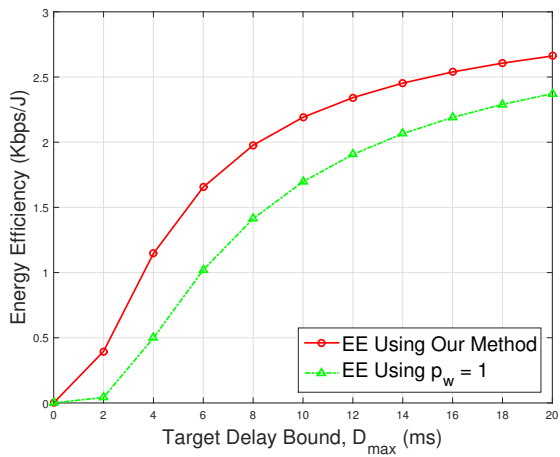}
\caption{Optimal energy efficiency under different delay-outage probability.}
\end{figure}

Fig. 4 shows a comparison of our proposed power control scheme and the scheme using the first method. The $x$-axis is the values of a possible delay bound while the $y$-axis is the cross-layer energy efficiency. Other parameters are set as follows: $\varepsilon=0.01$, $p=0.5$ and $\overline L=1488$ bit (an average Internet packet size found in~\cite{ref17}). It is shown that when the delay bound $D_{max}>2$ ms, the maximum energy efficiency improvement using our scheme can be 38.34\%. The reasons for this improvement are two-fold: 1) the power control mechanism based on the first method uses more than expected transmission power; 2) if $p_w = 1$, then $u^* = 0$ based on (\ref{eq:14_1}), which makes $\eta = \mu / (P_{tx} + P_c)$ based on (\ref{eq:5}). $\eta = \mu / (P_{tx} + P_c)$ is exactly the energy efficiency using a single-mode circuitry~\cite{ref12} and is apparently less energy efficient.

Fig. 5 shows the optimal energy efficiency under two target delay-outage constraints ($\{10 ms, 0.01\}$ and $\{100 ms, 0.01\}$) and different data arrival rates ($\overline L=1488$ bits, $p = \{0.1, 0.2,..., 1\}$). This figure shows an interesting result that the cross-layer energy efficiency versus the arrival rate is not a monotonic function but more like a unimodal function. In other words, there might exist an optimal data rate that achieves a maximum energy efficiency.

\begin{figure}[!t]
\centering
\includegraphics[width=2.9in]{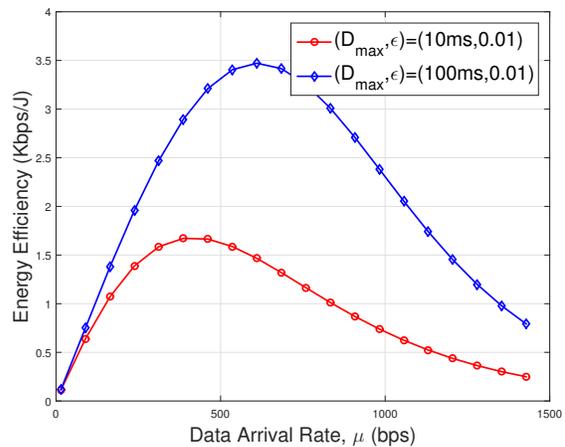}
\caption{Achieved energy efficiency under different arrival rates.}
\end{figure}

\section{Conclusion}

%The energy-efficient power control optimization under a predefined delay-outage probability were conventionally carried out by using only the QoS exponent (one of two effective capacity functions defined in the effective capacity model). However, it is unknown how the nonempty buffer probability (the other effective capacity function) affects the energy efficiency or the power control optimization problem.

It is important to design an energy-efficient power control mechanism under predefined delay-outage constraints in 5G wireless systems. Recent progress on this topic indicated the uses of a two-mode circuitry and the nonempty buffer probability measure improve a system's energy efficiency and QoS performance; however, no concrete work has emphasized their exact relations to power control problems. In this paper, we consider the use of a two-mode transceiver circuitry, and then propose an optimal power control scheme that maximizes a system's energy efficiency with precise delay-outage probability guarantee. Simulation results confirm the preciseness of our scheme, and the energy efficiency using our scheme is significantly higher than other schemes. Our results further indicate that the energy efficiency attains its maximum value at a specific arrival rate, which might help us design rate control and other resource allocation algorithms in future.

\section{Acknowledgment}

%The work was supported by National Nature Science Foundation of China Project under the Grant 61471058, Key National Science %Foundation of China under the Grant 61461136002, the Funds for Creative Research Groups of China under the Grant 61421061, Hong %Kong, Macao and Taiwan Science and Technology Cooperation Projects under the Grant 2016YFE0122900.

The work was supported by the National Nature Science Foundation of China Project under the Grants 61471058 \& 61231009, Key National Science Foundation of China under the Grant 61461136002, the National Science and Technology Major Project of China under Grant 2017ZX03001004, the Funds for Creative Research Groups of China under the Grant 61421061, Hong Kong, Macao and Taiwan Science and Technology Cooperation Projects under the Grant 2016YFE0122900, the Shenzhen Science and Technology Project (No. JSGG20150512153045135).

\appendices
\section{Proof of Proposition 1}

It has been proved in \cite{ref12} that under the conditions of Proposition 1, the probability that the queue length $Q$ exceeds a small backlog size bound $B$ can be approximated by
\begin{equation}
{\rm P}(Q > B) \approx {p_b}\exp ( - {u^*}B),
\label{eq:36}
%(36)
\end{equation}
where
\begin{equation}
{p_b} = {\rm P}(Q > 0) \approx 1 - {u^*}\overline L.
\label{eq:37}
%(37)
\end{equation}
and $p_b$ is termed the nonempty buffer probability.

In the same model, the steady-state delay $D$ is a random variable (RV) and its distribution satisfies \cite{ref14}
\begin{equation}
\mathop {\lim }\limits_{{t} \to \infty } \frac{1}{{{t}}}\log
{\rm{P}}\left( {D > {t}} \right) =  - {\theta ^*},
\label{eq:38}
%(38)
\end{equation}
where $t$ is a delay bound and ${\theta}^*$ is the delay exponent that is
\begin{equation}
{\theta ^*} = {u^*}{\alpha ^{\left( b \right)}}\left( {{u^*}} \right) =
\frac{1}{{{T_s}}}\log \left( {\frac{p}{{1 - {u^*}\overline L}} + 1 - p} \right).
\label{eq:39}
%(39)
\end{equation}
For small values of delay bound $t$, the CCDF of delay can be approximated by
\begin{equation}
\begin{aligned}
{\rm P}\left( {D > {t}} \right) &\approx {\rm P}\left( {D > 0} \right)\exp \left( { - {\theta
^{\rm{*}}}{t}} \right)\\
&={p_w}{\left( {\frac{{1 - {u^*}\overline L}}{{1 - {u^*}\overline L + p{u^*}\overline L}}} \right)^{\frac{t}{{{T_s}}}}},
\label{eq:40}
%(40)
\end{aligned}
\end{equation}
which is (\ref{eq:14}). $p_w$ in (\ref{eq:40}) is the nonzero delay probability. According to Little's Law, we have the following equation:
\begin{equation}
{\rm E}\left[ D \right] = \frac{{{\rm E}\left[ Q \right]}}{\mu},
\label{eq:41}
%(41)
\end{equation}
where $\mu$ is the average arrival rate as (\ref{eq:2}). From (\ref{eq:36}), ${\rm E}[Q] = \frac{{{p_b}}}{{{u^*}}}$; and from (\ref{eq:40}), ${\rm E}[D] = T_s\frac{{{p_w}\left( {p{u^*}\overline L - {u^*}\overline L + 1} \right)}}{{p{u^*}\overline L}}$. Therefore, the nonzero delay probability is
\begin{equation}
{p_w} = \frac{{{p_b}}}{{1 - {u^*}\overline L + p{u^*}\overline L}} = \frac{{1 - {u^*}\overline
L}}{{1 - {u^*}\overline L + p{u^*}\overline L}},
\label{eq:42}
%(42)
\end{equation}
which is (\ref{eq:14_1}).

\section{Proof of Lemma 1}

Based on the inequality: $\log_2\left(1+x\right)<x$, the right-hand side of (\ref{eq:18}) has a lower bound:
\begin{equation}
\!\!\!\begin{aligned}
&\int_0^\infty  {{e^{ - {u^*}{T_s}{B_c}\log_2(1 + \frac{{{P_{tx}}\gamma
}}{{{L_p}{N_0}{B_c}}})}}} {f_\gamma }\left( \gamma  \right)d\gamma  \\
&> \int_0^\infty  {{e^{ - {u^*}{T_s}{B_c}\frac{{{P_{tx}}\gamma
}}{{{L_p}{N_0}{B_c}}}}}} {f_\gamma }\left( \gamma  \right)d\gamma \\
&=\int_0^\infty  {\left( {\frac{m}{{\bar \gamma }}} \right)^m}{{e^{ -
\frac{{{u^*}{T_s}{P_{tx}}}}{{{L_p}{N_0}}}\gamma }}\frac{1}{{\Gamma (m)}}} {\gamma
^{m - 1}}{e^{ - \frac{m}{{\bar \gamma }}\gamma }}{\rm{ }}d\gamma.
\end{aligned}
\label{eq:44}
%(44)
\end{equation}
The right-hand side of (\ref{eq:44}) resembles a moment generating function of $\gamma$ and it can be further simplified:
\begin{equation}
\begin{aligned}
&\int_0^\infty {\left({\frac{m}{{\bar \gamma }}} \right)^m}{{e^{-
\frac{{{u^*}{T_s}{P_{tx}}}}{{{L_p}{N_0}}}\gamma }}\frac{1}{{\Gamma (m)}}}{\gamma
^{m-1}}{e^{-\frac{m}{{\overline \gamma }}\gamma }}{\rm{ }}d\gamma \\
&=\frac{1}{{{{\left({1+\frac{{mu^*{T_s}{P_{tx}}}}{{\overline\gamma{L_p}{N_0}}}}
\right)}^m}}}.
\end{aligned}
\label{eq:45}
%(45)
\end{equation}
By substituting the right-hand side of (\ref{eq:18}) with (\ref{eq:45}), we have a lower bound of the transmission power:
\begin{equation}
\!\!\begin{aligned}
&\frac{{1 - {u^*}\overline L }}{{1 - \left( {1 - p} \right){u^*}\overline L }} >
\frac{1}{{{{\left( {1 + \frac{{m{u^*}{T_s}{P_{tx}}}}{{\overline \gamma {L_p}{N_0}}}}
\right)}^m}}}\\
&\Rightarrow {P_{tx}}\!>\!\frac{{\overline \gamma {L_p}{N_0}}}{{m{T_s}}}\frac{1}{{{u^*}}}\left( \!{{{\left(\!{1\!+\!
\frac{{p{u^*}\overline L }}{{1\!-\!{u^*}\overline L }}} \right)}^{\frac{1}{m}}}\!-\!1} \!\right)\!=\!{P_l}.
\end{aligned}
\label{eq:46}
%(46)
\end{equation}

Since $P_{tx}>P_l$ and ${1 - {u^*}\bar L\left( {1 - p} \right)}>0$, the energy efficiency in \textbf{P2} has an upper bound:
\begin{equation}
\begin{aligned}
\eta  &=\!\frac{\mu }{{{P_{tx}}\left( {1\!-\!{u^*}\overline L\left( {1\!-\!p} \right)} \right) + {P_{idle}}{u^*}\overline L\left( {1 - p} \right) + {P_c}}} < \\
&\frac{\mu }{{{P_l}\left( {1 - {u^*}\overline L\left( {1 - p} \right)} \right) + {P_{idle}}{u^*}\overline L\left( {1 - p} \right) + {P_c}}} = {\eta _u}.
\label{eq:47}
%(47)
\end{aligned}
\end{equation}

%\begin{equation}
%\begin{aligned}
%{\eta _u} & = \frac{\mu }{{{P_l}{p_{tx}} + {P_{idle}}{p_{idle}} + {P_c}}} \\
%& > \frac{\mu }{{{P_{tx}}{p_{tx}} + {P_{idle}}{p_{idle}} + {P_c}}} = \eta.
%\end{aligned}
%\label{eq:47}
%%(47)
%\end{equation}

\section{Proof of Proposition 2}

Let $f(u^*)$ be the denominator of the energy efficiency $\eta{}$ in (\ref{eq:8}):
\begin{equation}
f\left( {{u^*}} \right) = {P_{tx}} + {P_c} - ({P_{tx}} - {P_{idle}}){u^*}\bar
L(1 - p).
\end{equation}
Based on Lemma 1, $f(u^*)$ can be re-expressed as
\begin{equation}
\!\!\begin{aligned}
f\left( {{u^*}} \right) &= {P_{tx}} + {P_c} - ({P_{tx}} - {P_{idle}}){u^*}\overline L(1 - p)\\
 &= \alpha {P_l} + {P_c} - (\alpha {P_l} - {P_{idle}}){u^*}\bar L(1 - p)\\
 &= \alpha {P_l}\left( {1\!-\!{u^*}\overline L(1\!-\!p)} \right)\!+\!{P_{idle}}{u^*}\overline L(1\!-\!p)\!+\!{P_c}\\
 &= \!\frac{{\bar \gamma {L_p}{N_0}}}{{m{T_s}}}\alpha \left(\!{{{\left(\!{1\!\!+\!\!\frac{{p{u^*}\overline L }}{{1\!\!-\!\!{u^*}\overline L }}} \right)}^{\frac{1}{m}}}\!\!-\!\!1} \right)\!\!\frac{{1\!-\!\left( \!{1\!-\!p} \right)\!\!{u^*}\overline L }}{{{u^*}}}\\
 &+ {P_{idle}}{u^*}\bar L(1 - p) + {P_c}\\
 &= c{\alpha}g\left( {{u^*}} \right) + h\left( {{u^*}} \right),
\end{aligned}
\end{equation}
where
\begin{equation}
\alpha(u^*)  = \frac{{{P_{tx}}}}{{{P_l}}},
\end{equation}
\begin{equation}
g\!( {{u^*}} \!)\!=\!\left(\!{{{\left( {1\!+\!\frac{{p{u^*}\overline L }}{{1\!-\!{u^*}\overline L }}} \right)}^{\frac{1}{m}}}\!-\!1} \right)\frac{{1\!-
\!{u^*}\overline L  + p{u^*}\overline L }}{{{u^*}}},
\label{eq:50}
%(50)
\end{equation}
\begin{equation}
c = {{\overline \gamma {L_p}{N_0}} \mathord{\left/
{\vphantom {{\overline \gamma {L_p}{N_0}} {m{T_s}}}} \right.
\kern-\nulldelimiterspace} {m{T_s}}}
\label{eq:51}
%(51)
\end{equation}
and
\begin{equation}
h\left( {{u^*}} \right) = {P_{idle}}{u^*}\overline L(1 - p) + {P_c}.
\label{eq:52}
%(52)
\end{equation}

$h\left(u^*\right)$ can be easily proved to be an increasing function of $u^*$ so the only tasks left us to do are to prove that $\alpha(u^*)$ and $g\left(u^*\right)$ are indeed increasing functions. $P_{tx}(u^*)$ and $P_l(u^*)$ have the relation below based on Lemma 1:
\begin{equation}
{\log _2}\left( {1 + \frac{{{P_{tx}}\gamma }}{{{L_p}{N_0}{B_c}}}} \right) = \frac{{{P_l}\gamma }}{{{L_p}{N_0}{B_c}}}.
\end{equation}
The ratio of $P_{tx}(u^*)$ to $P_l(u^*)$ is
\begin{equation}
\alpha(u^*)  = \frac{{{P_{tx}}}}{{{P_l}}} = \frac{{{2^{\frac{{{P_l}\gamma }}{{{L_p}{N_0}{B_c}}}}} - 1}}{{\frac{{{P_l}\gamma }}{{{L_p}{N_0}{B_c}}}}}.
\end{equation}
It is known that $\frac{{{2^x} - 1}}{x} \ge 1\left( {x \ge 0} \right)$ and $\frac{{{2^x} - 1}}{x}$ is an increasing function of $x$. Therefore, $\alpha(u^*)$ is an increasing function of $P_l$. Because $P_l$ is also an increasing function of $u^*$, $\alpha(u^*)$ is an increasing function of $u^*$.

%Let $x=\frac{pu^*\overline L}{1-u^*\overline L}$. $g(\!{{u^*}}\!)$ can be written as
%{\small
%\begin{equation}
%\begin{aligned}
%g(\!{x}\!) &=\!\left({{{\left(\!{1\!+\!\frac{{p{u^*}\overline L}}{{1\!-\!
%{u^*}\overline L }}} \right)}^{\frac{1}{m}}}\!-\!1} \right)\left( {p\overline L
%\frac{{1 - {u^*}\overline L }}{{p\overline L {u^*}}}\!+\!p\overline L } \right)\\
%&= p\overline L \left( {{{\left( {1 + x} \right)}^{\frac{1}{m}}} - 1}
%right)\left( {\frac{1}{x} + 1} \right).
%end{aligned}
%\label{eq:53}
%(53)
%\end{equation}
%}
Let $x = {\left( {\frac{{p{u^*}\overline L}}{{1 - {u^*}\overline L}} + 1} \right)^{\frac{1}{m}}}$, $g(\!{{u^*}}\!)$ can be further written as %$y={\left(1+x\right)}^{\frac{1}{m}}\Longleftrightarrow{}x=y^m-1$.
\begin{equation}
g\left( {{x}} \right) = p\overline L \frac{{\left( {x - 1}
\right){x^m}}}{{{x^m} - 1}}.
\label{eq:54}
%(54)
\end{equation}
Based on the result $(x^n-1)=\left(x-1\right)\left(x^{n-1}{+x}^{n-2}+x+1\right)$, (\ref{eq:54}) can be rewritten as
{\small
\begin{equation}
\begin{aligned}
g\left( {{x}} \right) &= p\overline L \frac{{\left( {x - 1}
\right){x^m}}}{{\left( {x - 1} \right)\left( {{x^{m - 1}} + {x^{m - 2}} +  \cdot
\cdot  \cdot  + x + 1} \right)}}\\
                        &= p\overline L \frac{1}{{\frac{1}{x} + \frac{1}{{{x^2}}} +  \cdot  \cdot  \cdot
 + \frac{1}{{{x^{m - 1}}}} + \frac{1}{{{x^m}}}}}.
\end{aligned}
\label{eq:55}
%(55)
\end{equation}
}

As $x$ is increasing with $u^*$ and $g\left(x\right)$ is increasing with $x$, $g\left(u^*\right)$ is an increasing function of $u^*$. Because $\alpha$, $h\left(u^*\right)$ and $g(u^*)$ are increasing functions, the $\eta_u$ is a decreasing function of $u^*$. Moreover, the transmission power $P_{tx}$ increases with $u^*$, so $\eta{}$ is a decreasing function of $P_{tx}$.

\section{Binary Search Algorithm to Find Transmission Power}

In this algorithm, ${\delta{}}_e={\alpha{}}^{(c)}-{\alpha{}}^{(b)}$ is a difference between the effective bandwidth and the effective capacity. ${\delta{}}_t$ is a precision tolerance, $P_s$ and $P_u$ are lower and upper bound on $P_{tx}$, respectively. We have two important observations for this algorithm:
\begin{enumerate}
	\item $P_{tx}$ should always be more than 0 and less than $P_{max}$;
	\item ${\delta{}}_e$ is a monotonically decreasing function because the effective bandwidth is a constant value and the effective capacity is a monotonically decreasing function.
\end{enumerate}

The following binary search algorithm is developed based on these two observations:
{\raggedright
\vspace{3pt} \noindent
\begin{tabular}{p{215pt}}
\hline
\parbox{248pt}{\raggedright
\textbf{Algorithm} finding the transmission power $P_{tx}$
} \\
\hline
\parbox{248pt}{\raggedright
\hspace{0.5em}1: Initialize  \{e.g.,${\delta{}}_t={10}^{-6}$ \}\\
\hspace{0.5em}2: $P_s=0$ \{a lower bound\}\\
\hspace{0.5em}3: $P_u\!=\!P_{max}$ \!\!\{an \!upper\! bound\! of\! the\! $1^{st}$ observation \\
that is sufficiently large\}\\
\hspace{0.5em}4: $P=(P_s+P_u)/2$ \{the $1^{st}$ guess about $P_{tx}$\}\\
\hspace{0.5em}5: ${\delta{}}_e={\alpha{}}^{(c)}-{\alpha{}}^{(b)}$\\
\hspace{0.5em}6: \textbf{while $\left\vert{}{\delta{}}_e\right\vert{}>{\delta{}}_t$ do}\\
\hspace{0.5em}7: \quad\textbf{if ${\delta{}}_t>0$ then}\\
\hspace{0.5em}8: \qquad$P_s=P$\\
\hspace{0.5em}9: \quad\textbf{else}\\
10: \qquad$P_u=P$\\
11: \quad\textbf{end if}\\
12: \quad$P=\left(P_s+\ P_u\right)/2$\\
13: \quad${\delta{}}_e={\alpha{}}^{(c)}-{\alpha{}}^{(b)}$\\
14: \textbf{end while}\\
15: $P_{tx}=P$\\
} \\
\hline
\end{tabular}
\vspace{2pt}
}

\end{document}